\documentclass[preprint, 12pt]{revtex4-1}

\usepackage{lineno,hyperref}
\usepackage{epsfig,amsmath}
\usepackage{subfigure}
\usepackage{graphicx}
\usepackage{dcolumn}
\usepackage{stmaryrd}
\usepackage{mathrsfs}
\usepackage{pifont}
\usepackage{amsthm}
\usepackage{amssymb}
\usepackage{bm}
\usepackage{latexsym}
\usepackage{color}
\usepackage{amsmath,graphicx,bbm,mathrsfs,amssymb,pst-all,bm,color}
\usepackage{mathrsfs}
\usepackage{algorithmic}
\usepackage[ruled]{algorithm}
\usepackage{multirow}
\usepackage{verbatim}
\usepackage{enumerate}
\usepackage{longtable}

\modulolinenumbers[5]
\begin{document}

\title{Quantum algorithm for PageRank computation through multistep quantum resonant transitions}
\author{Chuqing Wang$^1$}
\author{Hefeng Wang$^1$}
\email{wanghf@mail.xjtu.edu.cn}
\author{Hua Xiang$^2$}
\email{hxiang@whu.edu.cn}
\affiliation{$^{1}$School of Physics, Xi'an Jiaotong University, Xi'an,
710049, China}
\affiliation{$^{2}$School of Mathematics and Statistics, Wuhan University, Wuhan, 430072, China}

\begin{abstract}
We present a quantum algorithm for obtaining a quantum state that encodes
the PageRank vector of the Google matrix through multistep quantum
resonant transition~(mQRT). In the algorithm, the PageRank vector is encoded in the ground state of a problem Hamiltonian associated with the Google matrix. By dividing the web graph corresponding to the Google matrix into a series of subgraphs with nested structure, we construct a sequence of Hamiltonians based on the subgraphs to form a Hamiltonian evolution path from a simple initial Hamiltonian to the problem Hamiltonian. The ground state of the problem Hamiltonian is obtained by going through ground states of the intermediate Hamiltonians via QRT step by step. This algorithm requires only one ancillary qubit, and the runtime of the algorithm is proportional to the number of steps. It provides a new way for efficiently obtaining the quantum state of the PageRank vector of large-scale networks.
\end{abstract}

\maketitle

\section{Introduction}
The PageRank vector that provides webpage importance ranking was first
proposed by Brin and Page~\cite{brin1998anatomy}, it underlies the success
of the Google search engine. In the PageRank algorithm, webpages are
represented as nodes on a web graph, connected by directed edges that
represent links. The PageRank algorithm produces a node ranking score by
order of importance on a directed web graph, i.e., the PageRank vector,
which is the principal eigenvector of the adjacency matrix of the web graph~%
\cite{langville2006google}. The best available classical algebraic and
Markov Chain Monte Carlo~(MCMC) techniques used to evaluate the full
PageRank vector require a runtime that scales as $O(N)$ and $O\left[ N\log
\left( N\right) \right] $~\cite{VenegasAndraca2013b}, respectively, where $N$
is the number of webpages in the network, i.e. the size of the web graph.
Mathematically, the PageRank algorithm amounts to solving for the PageRank
vector of a sparse matrix at trillion-scale magnitude, which is expensive
for classical computers.

Quantum computation can outperform classical algorithms in solving some
problems with polynomial or exponential speed-up~\cite{nielsen2010quantum}.
In Ref.~\cite{VenegasAndraca2013b}, a quantum algorithm based on quantum
adiabatic evolution~(QAE)~\cite%
{Farhi2000,PowerLawScaling2013,VenegasAndraca2013} was proposed for
obtaining the PageRank vector of the Google matrix. In this algorithm, the
problem of computing the PageRank vector is transformed to finding the
ground state of a problem Hamiltonian that is constructed based on the
Google matrix. And by starting from a simple initial Hamiltonian and its ground state, the system is evolved adiabatically to the problem Hamiltonian and its ground state. Numerical simulations showed that runtime of the algorithm scales as $O\left[ \text{poly}\log (N)\right] $~\cite{VenegasAndraca2013b}. Further more, the PageRank problem was transformed into linear system of equations~%
\cite{ChapuisChkaiban2023} and solved by applying the HHL algorithm~\cite%
{Harrow2009,Childs2017}. The condition number $\kappa $ of the linear system
of equations for the PageRank problem is polynomial large~\cite%
{ChapuisChkaiban2023,Meyer2004}, therefore it can be solved efficiently.
Here, the term \textquotedblleft polynomially large\textquotedblright\
refers to a quantity $f(n)$ that is bounded both above and below by
polynomials in $n$. However, implementation of this algorithm requires large
number of ancillary qubits and the circuit of the algorithm is complicated
for a practical network.

One can also find in the literature some algorithms based on quantum walks that compute a PageRank-like centrality measure for graph nodes, known as quantum PageRank~\cite{Paparo2012}. In general, these quantum PageRank notions are not identical to the classical PageRank, nor do they yield the same ranking. Unlike algorithms that compute the PageRank vector of the Google matrix, those based on quantum random walks~\cite{kemp} have been used to compute the quantum PageRank of webpages~\cite%
{Paparo2012,Ambainis2003,pap1,pap2,Sanchez2012,Loke2017}. In these algorithms, the hyperlink relationships among webpages are mapped to quantum state transitions.

In this work, we present a quantum algorithm for obtaining a quantum state
that encodes the PageRank vector of the Google matrix through the multistep
quantum resonant transition~(mQRT)~\cite{whf2,Wang2021} approach. The
PageRank vector is encoded in the ground state of a problem Hamiltonian
based on the Google matrix. By dividing the web graph corresponding to the
Google matrix into a series of subgraphs with nested structure, we construct
a sequence of Hamiltonians based on the subgraphs to form a Hamiltonian
evolution path, which is from an initial Hamiltonian of a small
subgraph to the problem Hamiltonian of the web graph. The ground state of
the problem Hamiltonian is obtained by going through ground states of the
intermediate Hamiltonians via QRT step by step. The runtime of the algorithm is proportional to the number of steps of the algorithm, which is proportional to $\log N$ by applying a near bi-partitioning method to divide the web graph step by step. Compared to the QAE algorithm, our algorithm has the advantage that we only need to implement time-independent Hamiltonian
evolution. And compared to the algorithm based on the HHL algorithm, our
algorithm requires only one ancillary qubit and the complexity of the
circuit is reduced dramatically.

The structure of this work is as follows: In Sec.~II, we present the
algorithm for obtaining the quantum state encoding the PageRank vector of
the Google matrix; in Sec.~III, we provide an example to illustrate the
working procedures of the algorithm and perform numerical experiments on
practical networks to show the validity of the algorithm; we then close with
a discussion section.

\section{Quantum algorithm for computing the PageRank vector}

\subsection{A brief review of the mQRT approach}

In the mQRT approach for obtaining the ground state of a problem
Hamiltonian, first we construct a Hamiltonian evolution path from an initial
Hamiltonian to the problem Hamiltonian $H_{m}$ as $H_{0}\rightarrow
H_{1}\rightarrow \dots \rightarrow H_{m}$, then the system is evolved from
the ground state of the initial Hamiltonian to that of the problem
Hamiltonian by going through ground states of the intermediate Hamiltonians
sequentially via QRT in $m$ steps, i.e., $|\varphi ^{(0)}\rangle \rightarrow
|\varphi^{(1)}\rangle \rightarrow \dots \rightarrow |\varphi ^{(m)}\rangle $.

In the QRT method, a probe qubit is coupled to a quantum register that
represents the system. When the resonant transition condition is satisfied,
that is, the transition frequency of the probe qubit matches that of a
transition in the system, a quantum resonant transition occurs and the
system is evolved to its ground state by properly setting the parameters.
This process can be described by the Rabi-oscillation dynamics~\cite{cohen},
in which the probe qubit and the system exchange an excitation. The success
transition probability of the $k$th step of the approach is $p_{k}\approx
\sin ^{2}\left( ct_{k}d_{k}\right) $, where $c$ is the coupling coefficient,
$t_{k}$ is the evolution time, and $d_{k}=\left\vert \langle \varphi
^{(k-1)}|\varphi ^{(k)}\rangle \right\vert $ is the overlap between the
ground states of two adjacent Hamiltonians $H_{k-1}$ and $H_{k}$. The
transition probability scales as $O(1)$ when $t_{k}\sim 1/(cd_{k})$, and the
evolution of the step is efficient if $t_{k}$ is finite, which means that the
parameters $c$ and $d_{k}$ must not be exponentially small, i.e., lower
bounded by a polynomial function of the problem size. The coupling
coefficient $c$ is used to separate energy levels of the system, and it
determines the accuracy of the algorithm. It is set as $c<\Delta $ where $%
\Delta $ is the energy gap between the ground and the first excited states
of the Hamiltonian of the system.

The mQRT algorithm can be run efficiently provided: $i$) the energy gap between the ground and the first excited states of each Hamiltonian, and $ii$) the overlap between the ground states of any two adjacent Hamiltonians are not exponentially small~\cite{Wang2021}. In this work, we show that the first condition of the algorithm is always satisfied by constructing the intermediate Hamiltonians based on a series of Google matrices, and the second condition of the algorithm can be satisfied by applying some appropriate partitioning method for a practical web graph.

\subsection{Construction of the Hamiltonian evolution path}

The entire internet can be modeled as a directed graph $D\left( V,E\right) $%
, where the vertex set $V$ represents webpages and the edge set $E$ denotes
links between webpages. By applying the mQRT algorithm, the entire web graph
is partitioned into a series of subgraphs $D=D_{m},$ $D_{m-1},\ldots
,D_{1},D_{0}$ that form a nested structure $D_{m}\supset D_{m-1}\supset
\cdots \supset D_{1}\supset D_{0}$. In each step, a graph is bi-partitioned
in a relatively balanced way considering the distribution of the nodes and
the edges, such that for two adjacent subgraphs $D_{k-1}$ and $D_{k}$ ($%
k=1,2,\ldots ,m$), the portion of $D_{k-1}$ in $D_{k}$ is not exponentially
small, so that the overlap between the PageRank vectors of the corresponding
Google matrices is not exponentially small. In general, such partitioning
requires some knowledge of the structure of the network in advance, i.e.,
distribution of the nodes and edges in the web graph. The number of
subgraphs $m$ is in the order of $O(\log N)$ under this condition.

There are networks with different structures, e.g., social networks, communication networks, citation networks, web graphs, etc. By utilizing the inherent properties of webpages, we can divide the web graph following the principle of cutting fewer edges without destroying the whole structure of the graph too much. For example, for a graph whose related webpages distribute
relatively evenly worldwide, the graph of webpages can be regarded as a
world map, and we can partition the web graph geographically. E.g., in the
first step, the graph of webpages of Asia is taken as a subgraph of the
whole graph of the webpages; in the second step, the graph of webpages of a
country that takes polynomial large portion of Asia is taken as a subgraph
of Asia; in the third step, we can take the webpages of a province of the
country as a subgraph, and we can continue this partition process to select
a city of the province, then a town of the city, and so on. This partition
method takes into account geographical structure of the network to ensure
that each subgraph is a polynomial large portion of the previous one. Then
the overlap between the PageRank vectors of two Google matrices associated
with a pair of adjacent graphs is not exponentially small, i.e., lower
bounded by a polynomial large number, e.g. $0.01$. Other partition methods
such as multi-level strategy can also be used. For networks with different
structures, we can find a way of partitioning the web graph in practice by
exploring the internal structure of the network. We show some practical
examples of partitioning the web graph in the next section.

In the following, we describe construction of the Hamiltonian evolution path
in the mQRT algorithm for obtaining PageRank vector of the Google matrix.
For the graph $D_{k}$ ($k=0,1,\ldots ,m$), the corresponding adjacency
matrix $S^{(k)}$ is of dimension $N_{k}$, the entries of the matrix $S^{(k)}$
are defined as:
\begin{equation}
S_{ij}^{(k)}=%
\begin{cases}
\frac{1}{\text{outdegree}(i)} & \text{if node }j\text{ links to node }i, \\
\frac{1}{N_{k}} & \text{if node }i\text{ is a dangling node (no outgoing
links)}, \\
0 & \text{otherwise.}%
\end{cases}%
\end{equation}%
This ensures that even in the case of dangling nodes, random walks can
continue. Here the adjacency matrix $S^{(k)}$ is row-stochastic. The
corresponding Google matrix $G_{k}$ is defined as:
\begin{equation}
G_{k}=\alpha S^{(k)}+\frac{(1-\alpha )}{N_{k}}J_{k},
\end{equation}%
where $J_{k}$ is an all-one matrix of dimension $N_{k}$, and $0<\alpha <1$
is a damping factor and is usually set as $\alpha =0.85$~\cite{Meyer2004}.
The largest eigenvalue of $G_{k}$ is one, and the corresponding eigenvector $%
\pi _{k}$ is the PageRank vector~\cite{langville2006google}, they satisfy
the equation
\begin{equation}
\pi _{k}^{T}G_{k}=\pi _{k}^{T}.
\end{equation}%
The Hamiltonian evolution path for the mQRT algorithm is constructed as
\begin{equation}
H_{k}=(I_{k}-G_{k})(I_{k}-G_{k})^{\dagger }+I_{k},\text{ \ \ }k=0,1,\ldots
,m.
\end{equation}%
where $I_{k}$ is the identity operator of dimension $N_{k}$. The dimension of the Hilbert space of the intermediate Hamiltonians $H_{k}$ depends on how
the original graph is divided step by step. In the examples of the next
section, the graphs are bi-partitioned in each step. The original graph has $N$ nodes, and the dimension of the Hilbert space is $N$, here we assume that $N$ is in a power of two, the nodes are
encoded in the computation basis state~(CBS) $\{|0\rangle ,\ldots
,|N-1\rangle \}$, the problem Hamiltonian $H_{m}$ is constructed in this
basis set. Then the Hamiltonian $H_{m-1}$ is constructed in CBS $\{|0\rangle
,\ldots ,|N/2-1\rangle \}$, and $H_{m-2}$ is constructed in CBS $\{|0\rangle
,\ldots ,|N/2^{2}-1\rangle \}$,$\ldots $, etc.

In the algorithm, when two adjacent Hamiltonians $H_{k-1}$ and $H_{k}$ are called, the Hamiltonian $H_{k-1}$ in the small Hilbert space of dimension $N_{k-1}$ can be embedded in a larger Hilbert space of dimension $N_{k}$. Note in Eq.~($4$), we add a term $I_{k}$ so that the ground state eigenvalue of each Hamiltonian is $1$, this allows us to set the parameters of the algorithm to satisfy the resonant transition condition as shown below. The ground state $|\pi _{0}^{\left( k\right) }\rangle $ of $H_{k}$ encodes the PageRank vector $\pi _{k}$ of the Google matrix $G_{k}$ of the graph $D_{k}$. The energy gap between the ground and the first excited states of $H_{k}$ is $\Delta \geq(1-\alpha )^{2}$, which is not exponentially small, therefore the first condition of the mQRT algorithm is always satisfied. In this algorithm, we are to find the ground state of $H_{m}$ by starting from the ground state of $H_{0}$ via the mQRT algorithm in $m$ steps.

\subsection{Procedures of the algorithm}

The algorithm requires ($n+1$) qubits with a probe qubit and an $n$-qubit
quantum register $R$ representing a system of dimension $N=2^{n}$. In the $k$%
th step~($k=1,\ldots ,m$) of the algorithm, we are to prepare the ground
state $|\pi _{0}^{\left( k\right) }\rangle $ of $H_{k}$, given the
Hamiltonian $H_{k}$, $H_{k-1}$ and its ground state $|\pi _{0}^{\left(
k-1\right) }\rangle $, the ground state eigenvalues of both $H_{k-1}$ and $%
H_{k}$ are $E_{0}^{\left( k-1\right) }=E_{0}^{\left( k\right) }=1$. The
algorithm Hamiltonian of the $k$th step is
\begin{equation}
H^{(k)}=-\frac{1}{2}\omega \sigma _{z}\otimes I_{k}+H_{R}^{(k)}+c\sigma
_{x}\otimes I_{k},
\end{equation}%
where%
\begin{equation}
H_{R}^{(k)}=\beta |1\rangle \langle 1|\otimes H_{k-1}+|0\rangle \langle
0|\otimes H_{k}
\end{equation}%
and $\beta $ is used to rescale the eigenvalues of the Hamiltonian $H_{k-1}$%
, $\omega $ is the frequency of the probe qubit, and $c$ is the coupling
coefficient. The parameter $\beta $ is set such that $E_0^{(k)}-\beta
E_0^{(k-1)}=\omega $, which satisfies the condition of resonant transition
between the probe qubit and the transition between states $|\pi _{0}^{\left(
k-1\right) }\rangle $ and $|\pi _{0}^{\left( k\right) }\rangle $. Here, $%
E_0^{(k-1)}=E_0^{(k)}=1$, we set $\omega =2$ and $\beta =-1$ to satisfy the
resonant condition, the evolution time is $t_{k}$. Procedures of the $k$th
step of the algorithm are as follows:

$i$) Set the probe qubit in its excited state $|1\rangle $ and the register $%
R$ in state $|\pi _{0}^{\left( k-1\right) }\rangle $;

$ii$) Implement time evolution operator $U_{k}=\exp \left[ -iH^{\left(
k\right) }t_{k}\right] $;

$iii$) Measure the probe qubit in its computational basis.

After applying the operation $U_{k}$, the system is approximately in an
entangled state $\sqrt{1-p_{k}}|1\rangle |\pi _{0}^{\left( k-1\right)
}\rangle +\sqrt{p_{k}}|0\rangle |\pi _{0}^{\left( k\right) }\rangle $, where
$p_{k}\approx \sin ^{2}\left( ct_{k}d_{k}\right) $ is decay probability of
the probe qubit of the $k$th step, and $d_{k}=\left\vert \langle \pi
_{0}^{\left( k-1\right) }|\pi _{0}^{\left( k\right) }\rangle \right\vert $
is the overlap between the ground states of the Hamiltonians $H_{k-1}$ and $%
H_{k}$. If the measurement outcome is in state $|0\rangle $, it indicates
that the desired state $|\pi _{0}^{\left( k\right) }\rangle $ is obtained on
the register $R$, and we proceed to the next step; otherwise if the
measurement outcome is in state $|1\rangle $, it means that the register $R$
remains in the state $|\pi _{0}^{\left( k-1\right) }\rangle $, then we
repeat procedures $ii$)-$iii$) until the measurement outcome on the probe
qubit is in state $|0\rangle $. Detailed derivation of the evolution of the $k$th step of the algorithm is as follows.

\subsection{Derivation of the evolution dynamics of the algorithm}

The Hamiltonians $H_{k-1}$ and $H_{k}$ satisfy $H_{k-1}|\pi _{j}^{\left(
k-1\right) }\rangle =E_{j}^{\left( k-1\right) }|\pi _{j}^{\left( k-1\right)
}\rangle $, and $H_{k}|\pi _{j}^{\left( k\right) }\rangle =E_{j}^{\left(
k\right) }|\pi _{j}^{\left( k\right) }\rangle $, respectively, where $|\pi
_{j}^{\left( k-1\right) }\rangle $ and$\ |\pi _{j}^{\left( k\right) }\rangle
\ $are the $j$th eigenstates with eigenvalues $E_{j}^{\left( k-1\right) }$
and $E_{j}^{\left( k\right) }$ of $H_{k-1}$ and $H_{k}$, respectively. Let
\begin{equation}
H_{0}^{\left( k\right) }=-\frac{1}{2}\omega \sigma _{z}\otimes
I_{k}+H_{R}^{\left( k\right) },
\end{equation}%
then the algorithm Hamiltonian $H^{\left( k\right) }$ of Eq.~($5$) can be
written as
\begin{equation}
H^{\left( k\right) }=H_{0}^{\left( k\right) }+cW^{(k)},
\end{equation}%
where $W^{(k)}=\sigma _{x}\otimes I_{k}$. The Hamiltonian $H_{0}^{\left(
k\right) }$ has eigenstates%
\begin{equation}
H_{0}^{\left( k\right) }|1\rangle |\pi _{j}^{\left( k-1\right) }\rangle
=\left( \frac{\omega }{2}\!+\!\beta E_{j}^{(k-1)}\right) |1\rangle |\pi
_{j}^{\left( k-1\right) }\rangle ,
\end{equation}%
and%
\begin{equation}
H_{0}^{\left( k\right) }|0\rangle |\pi _{j}^{\left( k\right) }\rangle
=\left( \frac{-\omega }{2}\!+\!E_{j}^{(k)}\right) |0\rangle |\pi
_{j}^{\left( k\right) }\rangle .
\end{equation}%
When the resonance condition is satisfied, the system is evolved from
initial state $|1\rangle |\pi _{0}^{\left( k-1\right) }\rangle $ to the
state $|0\rangle |\pi _{0}^{\left( k\right) }\rangle $.

For a probe qubit coupled to a two-level system described by the Hamiltonian in Eq.~($5$), according to the Rabi's formula~\cite{cohen}, the maximum
transition probability from the ground state to the excited state of the
two-level system becomes higher as the transition frequency between the
two-level system gets closer to the frequency of the probe qubit. Based on
this observation, the upper bound of the error of the $k$th step, that is,
the upper bound of the transition probability from the initial state $|\pi
_{0}^{\left( k-1\right) }\rangle $ to the excited states of the Hamiltonian $H_k$, can be obtained by assuming all other eigenstates are degenerate at
the first excited state $|\pi _{1}^{\left( k\right) }\rangle $ of $H_{k}$
that is closest to the state $|\pi _{0}^{\left( k\right) }\rangle $. The
states $|\pi _{1}^{\left( k-1\right) }\rangle $ and $|\pi
_{1}^{\left(k\right) }\rangle $ are the first excited states of the
Hamiltonians $H_{k-1}$ and $H_{k}$, respectively, and $|\pi _{1}^{\left(
k-1\right) }\rangle $ is orthogonal to $|\pi _{0}^{\left( k-1\right)
}\rangle $, and $|\pi_{1}^{\left( k\right) }\rangle $ is orthogonal to $|\pi
_{0}^{\left( k\right) }\rangle $. Then in basis $\left\{ |1\rangle |\pi
_{0}^{\left( k-1\right) }\rangle ,|1\rangle |\pi _{1}^{\left( k-1\right)
}\rangle ,|0\rangle |\pi _{0}^{\left( k\right) }\rangle ,|0\rangle |\pi
_{1}^{\left( k\right) }\rangle \right\} $, the Hamiltonian ${H^{(k)}}$ can
be written as:
\begin{equation}
{H^{(k)}=}\left(
\begin{array}{cccc}
E_{0} & 0 & c{d}_{k} & c\sqrt{1-{d}_{k}^{2}} \\
0 & E_{2} & c\sqrt{1-{d}_{k}^{2}} & c{d}_{k} \\
c{d}_{k} & c\sqrt{1-{d}_{k}^{2}} & E_{1} & 0 \\
c\sqrt{1-{d}_{k}^{2}} & c{d}_{k} & 0 & E_{3}%
\end{array}%
\right) ,
\end{equation}%
where $E_{0}=\frac{\omega }{2}+\beta E_{0}^{(k-1)}=\frac{-\omega }{2}%
+E_{0}^{(k)}=E_{1}$ under the resonance condition, and $E_{2}=E_{0}+\Delta =
E_{3} $, we already have $\Delta \geq (1-\alpha )^{2}>c$, and $%
d_{k}=|\langle \pi _{0}^{\left( k-1\right) }|\pi _{0}^{\left( k\right)
}\rangle |$. Then we have
\begin{equation}
H_{0}^{\left( k\right) }{=}\left(
\begin{array}{cccc}
E_{0} & 0 & 0 & 0 \\
0 & E_{2} & 0 & 0 \\
0 & 0 & E_{0} & 0 \\
0 & 0 & 0 & E_{2}%
\end{array}%
\right) ,
\end{equation}%
and
\begin{equation}
W{^{(k)}=}\left(
\begin{array}{cccc}
0 & 0 & {d}_{k} & \sqrt{1-{d}_{k}^{2}} \\
0 & 0 & \sqrt{1-{d}_{k}^{2}} & {d}_{k} \\
{d}_{k} & \sqrt{1-{d}_{k}^{2}} & 0 & {0} \\
\sqrt{1-{d}_{k}^{2}} & {d}_{k} & 0 & 0%
\end{array}%
\right) .
\end{equation}%

Let $|\psi \left( t\right) \rangle =c_{0}(t)e^{-iE_{0}t}|1\rangle |\pi
_{0}^{\left( k-1\right) }\rangle +c_{1}(t)e^{-iE_{2}t}|1\rangle |\pi
_{1}^{\left( k-1\right) }\rangle +c_{2}(t)e^{-iE_{0}t}|0\rangle |\pi
_{0}^{\left( k\right) }\rangle +c_{3}(t)e^{-iE_{3}t}|0\rangle |\pi
_{1}^{\left( k\right) }\rangle $.  Solving the Schr\"{o}dinger equation, we
have
\begin{equation}
i\frac{d}{dt}c_{j}(t)=c\sum_{l}e^{i(E_{j}-E_{l})t}W_{jl}^{\left( k\right)
}c_{l}(t),
\end{equation}%
that is,
\begin{eqnarray}
i\frac{d}{dt}c_{0}(t)\! &=&c\!\left[ \!{d}_{k}c_{2}(t)\!+\!e^{-i\Delta t}%
\sqrt{1\!-\!{d}_{k}^{2}}c_{3}(t)\!\right],  \notag \\
i\frac{d}{dt}c_{1}(t)\! &=&c\!\left[ \!e^{-i\Delta t}\sqrt{1\!-\!{d}_{k}^{2}}%
c_{2}(t)\!+\!e^{-i\Delta t}{d}_{k}c_{3}(t)\!\right],  \notag \\
i\frac{d}{dt}c_{2}(t)\! &=&c\!\left[ \!{d}_{k}c_{0}(t)\!+\!e^{i\Delta t}%
\sqrt{1\!-\!{d}_{k}^{2}}c_{1}(t)\!\right],  \notag \\
i\frac{d}{dt}c_{3}(t)\! &=&c\!\left[ \!e^{i\Delta t}\sqrt{1\!-\!{d}_{k}^{2}}%
c_{0}(t)\!+\!{d}_{k}c_{1}(t)\!\right] .
\end{eqnarray}%
By setting the initial state as $|1\rangle |\pi _{0}^{\left( k-1\right)
}\rangle $, we have $c_{0}(0)=1$ and $c_{j}(0)=0$ for $j\neq 0$. The
coefficients of $c_{0}(t)$ and $c_{2}(t)$ are proportional to one, so they
oscillate slowly in time under the resonance condition, while the
coefficients $c_{1}(t)$ and $c_{3}(t)$ oscillate much more rapidly. By
applying the secular approximation~\cite{cohen1}, the rapid oscillating
terms are neglected since their contribution is negligible when integrated
over time. Then the above equations are simplified to
\begin{eqnarray}
i\frac{d}{dt}c_{0}(t) &=&c{d}_{k}c_{2}(t),  \notag \\
i\frac{d}{dt}c_{2}(t) &=&c{d}_{k}c_{0}(t).
\end{eqnarray}%
Under the secular approximation, the algorithm Hamiltonian in the QRT method
can be spanned in basis $\left\{ |1\rangle |\pi _{0}^{\left( k-1\right)
}\rangle ,|0\rangle |\pi _{0}^{\left( k\right) }\rangle \right\} $ in form of
\begin{equation}
{H^{(k)\prime }=}\left(
\begin{array}{cc}
E_{0}^{(k)}-\frac{1}{2}\omega & c{d_{k}} \\
c{d}_{k} & E_{0}^{(k)}-\frac{1}{2}\omega%
\end{array}%
\right) .
\end{equation}%
The corresponding time evolution operator is
\begin{equation}
U_{k}^{\prime }=\exp \left[ -i{H^{(k)\prime }}t_{k}\right] {=}\left(
\begin{array}{cc}
\cos \theta _{k} & -i\sin \theta _{k} \\
-i\sin \theta _{k} & \cos \theta _{k}%
\end{array}%
\right) ,
\end{equation}%
where $\theta _{k}=c{d}_{k}t_{k}$. The unitary operator $U_{k}^{\prime }$
transforms the initial state $|1\rangle |\pi _{0}^{\left( k-1\right)
}\rangle $ to an entangled state
\begin{equation}
U_{k}^{\prime }|1\rangle |\pi _{0}^{\left( k-1\right) }\rangle =\cos \theta
_{k}|1\rangle |\pi _{0}^{\left( k-1\right) }\rangle -i\sin \theta
_{k}|0\rangle |\pi _{0}^{\left( k\right) }\rangle .
\end{equation}%
By ignoring the phase factor, the above entangled state can be written as $%
\sqrt{1-p_{k}}|1\rangle |\pi _{0}^{\left( k-1\right) }\rangle +\sqrt{p_{k}}%
|0\rangle |\pi _{0}^{\left( k\right) }\rangle $, where the probability of
the initial state $|1\rangle |\pi _{0}^{\left( k-1\right) }\rangle $ being
evolved to the state $|0\rangle |\pi _{0}^{\left( k\right) }\rangle $ is $%
p_{k}\approx \sin ^{2}\left( ct_{k}d_{k}\right) $ under the secular
approximation. Considering errors accumulated in each step of the algorithm,
in which the register is evolved to the excited states of an intermediate
Hamiltonian, the success probability of the algorithm satisfies $P_{\text{%
succ}}>1/e$ in the asymptotic limit of $m$~\cite{Wang2021}.

The evolution time $t_{k}$ of the $k$th step satisfies $t_{k}\sim 1/(cd_{k})$%
, such that the success probability of the step is polynomial large. The
parameter $c<\Delta $, and $\Delta \geq (1-\alpha )^{2}$. By applying an
appropriate partitioning method for the graph $D_{k}$, the overlap $d_{k}$
can be polynomial large, then the evolution time $t_{k}$ is finite.
Therefore the runtime of the algorithm for obtaining the state $|\pi
_{0}^{\left( m\right) }\rangle $ that encodes the PageRank vector of the
Google matrix scales as $O\left( \sum_{k=1}^{m}\frac{1}{(1-\alpha )^{2}d_{k}}%
\right) $.

\subsection{Implementation of the algorithm}

In the following, we discuss implementation of the unitary operator $%
U_{k}(t)=\exp \left( -iH^{\left( k\right) }t_{k}\right) $ of the algorithm,
which is a time-independent Hamiltonian simulation operation. In Hamiltonian
simulation, we need to construct a sequence of elementary quantum gates to
approximate the time evolution of a Hamiltonian. The scaling of the
simulation complexity is estimated as either the number of quantum gates or
the number of times the Hamiltonian is queried. There are two most used
methods for Hamiltonian simulation, the product formula method and the
quantum signal processing~(QSP)~\cite{QSP,low2} method, and we estimate the
complexity of simulating the algorithm Hamiltonians using these methods.

The product formula method using the first order Lie-Trotter-Suzuki
approximation~\cite{Trotter}, and the time evolution of the Hamiltonian is
in form of $U(t)=e^{-iHt}=\left(
\prod\nolimits_{j=1}^{p}e^{-ih_{j}t/M}\right) ^{M}\!+\!O\left(
t^{2}/M\right) $, where the number of Trotter steps $M=O(t^{2}/\epsilon )$
within an error $\epsilon $. Each segment $e^{-ih_{j}t/M}$ can be
implemented directly, the unitary $U(t)$ can be approximated within error $%
\epsilon $ by a sequence of $O(p\left\Vert H\right\Vert ^{2}t^{2}/\epsilon )$
elementary quantum gates~\cite{loyd,childsth}, where $\left\Vert
H\right\Vert $ is the norm of $H$ and $p$ is the number of local interacting
terms $h_{j}$ in the Hamiltonian. The algorithm based on QSP achieves the
optimal scaling of $O(s\left\Vert H\right\Vert _{\max }t+\frac{\log
1/\epsilon }{\log \log 1/\epsilon })$ for Hamiltonian simulation. For a
practical network, a webpage is usually connected to a limited number of
webpages, we can assume that the Hamiltonian matrices in the algorithm are
sparse. As shown in Sec.~II, the runtime of the algorithm scales as $%
O(\sum_{k=1}^{m}t_{k})$, where $t_{k}\thicksim \frac{1}{cd_{k}}$ is the
runtime of the $k$th step of the algorithm. In our algorithm, by applying
the product formula method, the time evolution of the algorithm Hamiltonian
in Eqs.~($5-6$) of the $k$th step is in form of%
\begin{eqnarray}
U_{k}(t) &=&e^{-iH^{\left( k\right) }t_{k}}  \notag \\
&=&\left(e^{-i\sigma _{z}t_{k}/M}e^{i|1\rangle \langle 1|\otimes H_{k-1}t_{k}/M}e^{-i|0\rangle \langle 0|\otimes
H_{k}t_{k}/M}e^{-ic\sigma _{x}t_{k}/M}\right)^{M}+O\left(t_{k}^{2}/M\right).
\end{eqnarray}%
Then our algorithm scales as $O(\sum_{k=1}^{m}$poly$(n,s)(\left\Vert
H^{(k)}\right\Vert ^{2}t_{k}^{2}/\epsilon )$ in both query complexity and
gate complexity using the product formula method in the first order
Lie-Trotter-Suzuki approximation. By applying the QSP algorithm, the query
complexity of our algorithm scales as $O\left( \sum_{k=1}^{m}s\left\Vert
H^{(k)}\right\Vert _{\max }t_{k}+\frac{\log 1/\epsilon }{\log \log
1/\epsilon }\right) $, and the gate complexity scales as $O\left(
\sum_{k=1}^{m}\text{poly}(n,s)\left( s\left\Vert H^{(k)}\right\Vert _{\max
}t_{k}+\frac{\log 1/\epsilon }{\log \log 1/\epsilon }\right) \right) $.

\section{Numerical simulation of the algorithm}

In this section, we first use a graph of $16$ webpages as an example to
illustrate how to obtain the quantum state that encodes the PageRank vector
of the Google matrix through the mQRT algorithm, then we present two
examples of practical network to show the robustness of the algorithm.

\subsection{A simple small graph}

The graph of $16$ webpages is shown in Fig.~$1$. The adjacency matrix $S$ of
the graph can be constructed as%
\begin{equation}
S={%
\begin{pmatrix}
0 & \frac{1}{2} & \frac{1}{2} & 0 & 0 & 0 & 0 & 0 & 0 & 0 & 0 & 0 & 0 & 0 & 0 & 0 \\
0 & 0 & \frac{1}{2} & 0 & 0 & 0 & \frac{1}{2} & 0 & 0 & 0 & 0 & 0 & 0 & 0 & 0 & 0 \\
\frac{1}{2} & 0 & 0 & \frac{1}{2} & 0 & 0 & 0 & 0 & 0 & 0 & 0 & 0 & 0 & 0 & 0 & 0 \\
1 & 0 & 0 & 0 & 0 & 0 & 0 & 0 & 0 & 0 & 0 & 0 & 0 & 0 & 0 & 0 \\
0 & 0 & 0 & 0 & 0 & \frac{1}{2} & 0 & \frac{1}{2} & 0 & 0 & 0 & 0 & 0 & 0 & 0 & 0 \\
0 & 0 & 0 & 0 & 0 & 0 & 0 & 1 & 0 & 0 & 0 & 0 & 0 & 0 & 0 & 0 \\
0 & 0 & 0 & 0 & 0 & 0 & 0 & 0 & \frac{1}{2} & \frac{1}{2} & 0 & 0 & 0 & 0 & 0 & 0 \\
0 & 0 & 0 & 0 & 0 & 0 & 1 & 0 & 0 & 0 & 0 & 0 & 0 & 0 & 0 & 0 \\
0 & 0 & \frac{1}{3} & 0 & 0 & 0 & 0 & 0 & 0 & \frac{1}{3} & \frac{1}{3} & 0 & 0 & 0 & 0 & 0 \\
\frac{1}{16} & \frac{1}{16} & \frac{1}{16} & \frac{1}{16} & \frac{1}{16} & \frac{1}{16} & \frac{1}{16} & \frac{1%
}{16} & \frac{1}{16} & \frac{1}{16} & \frac{1}{16} & \frac{1}{16} & \frac{1}{16} & \frac{1}{16} & \frac{1}{16} & \frac{1}{16} \\
0 & 0 & 0 & 0 & 0 & 0 & 0 & 0 & \frac{1}{3} & \frac{1}{3} & 0 & \frac{1}{3} & 0 & 0 & 0 & 0 \\
\frac{1}{16} & \frac{1}{16} & \frac{1}{16} & \frac{1}{16} & \frac{1}{16} & \frac{1}{16} & \frac{1}{16} & \frac{1%
}{16} & \frac{1}{16} & \frac{1}{16} & \frac{1}{16} & \frac{1}{16} & \frac{1}{16} & \frac{1}{16} & \frac{1}{16} & \frac{1}{16} \\
0 & 0 & \frac{1}{2} & 0 & 0 & 0 & 0 & 0 & 0 & 0 & 0 & 0 & 0 & 0 & \frac{1}{2} & 0 \\
0 & 0 & \frac{1}{3} & 0 & 0 & 0 & 0 & 0 & 0 & 0 & 0 & 0 & 0 & 0 & \frac{1}{3} & \frac{1}{3} \\
0 & 0 & 0 & 0 & 0 & 0 & 0 & 0 & 0 & 0 & 0 & 0 & 0 & \frac{1}{2} & 0 & \frac{1}{2} \\
0 & 0 & 0 & 0 & 0 & 0 & 0 & 0 & 0 & 0 & 0 & 0 & 0 & 1 & 0 & 0%
\end{pmatrix}.}
\end{equation}%
\begin{figure}[tbp]
\includegraphics[width=0.65\linewidth]{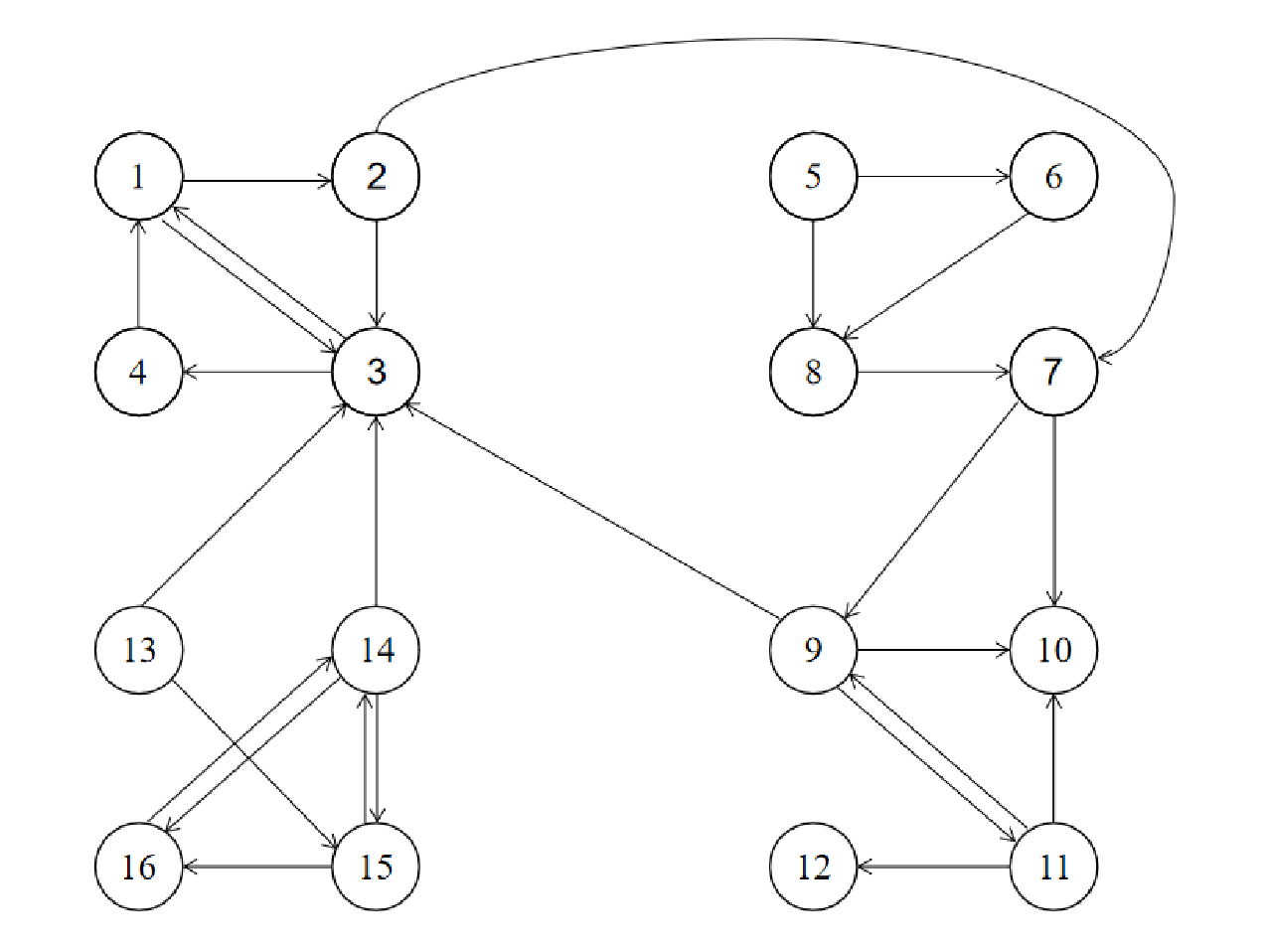}
\caption{Directed graph of $16$ webpages}
\label{fig1}
\end{figure}
\begin{figure}[tbp]
\includegraphics[width=0.65\linewidth]{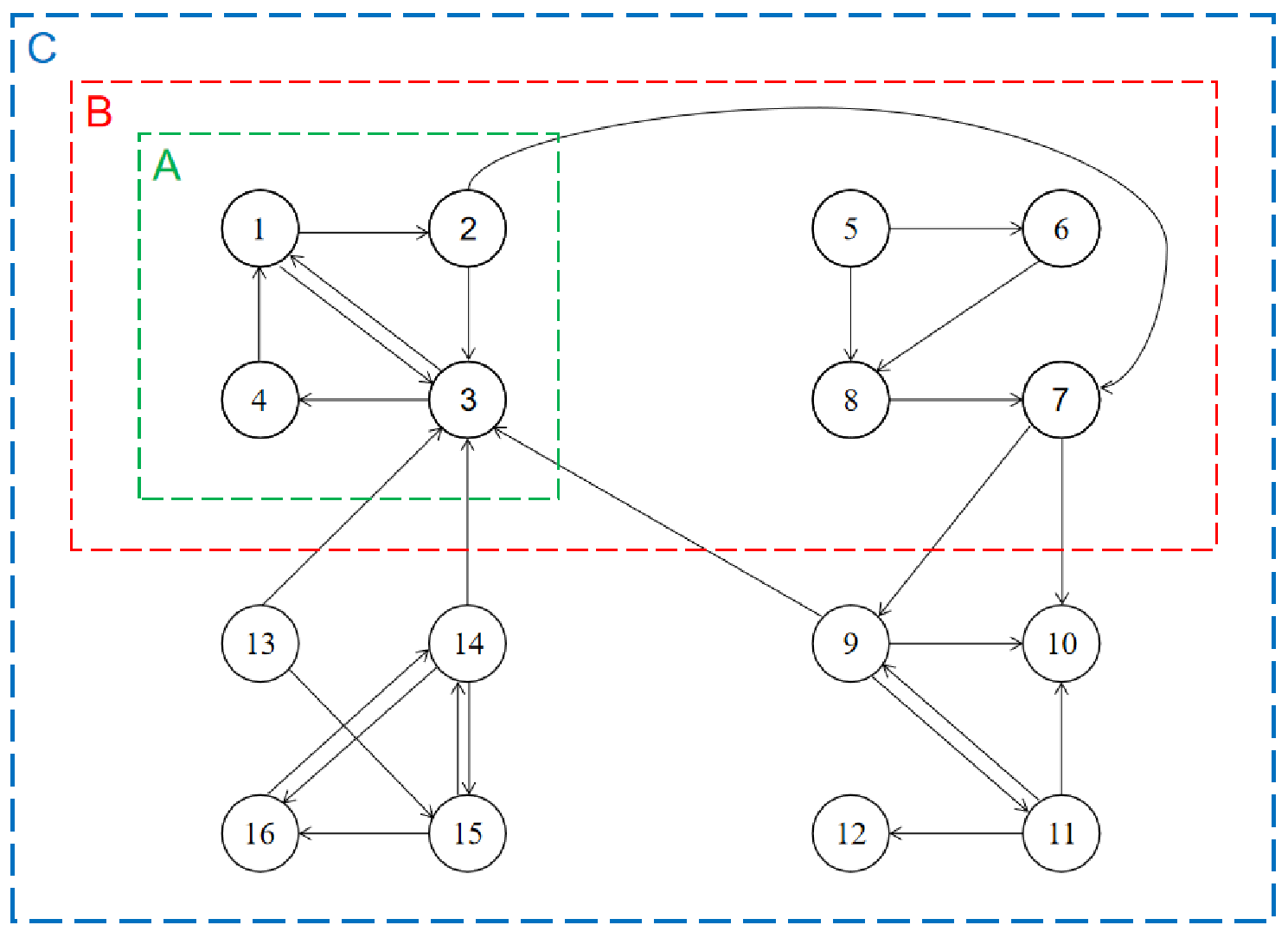}
\caption{Partition of the graph in Fig.~$1$. The whole graph $C$ is in the
blue dashed square with $16$ nodes; subgraph $B$ is in the red dashed square
and has $8$ nodes; subgraph $A$ is in the green dashed square with $4$
nodes. }
\label{fig2}
\end{figure}
\begin{figure}[tbp]
\includegraphics[width=1\linewidth]{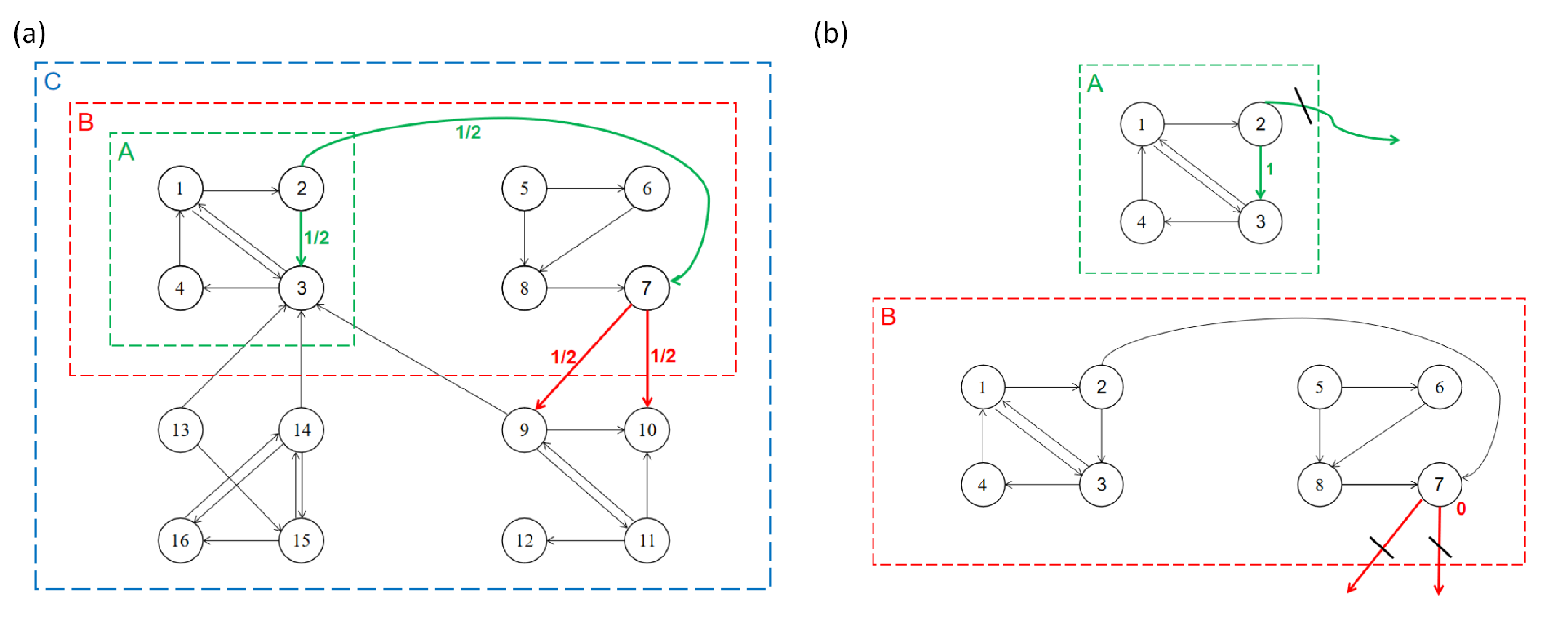}
\caption{Weight redistribution of edges in the subgraphs. (a) Weight
distribution of nodes $2$ and $7$ before partition. (b) Weight
redistribution of nodes $2$ and $7$ after partition.}
\label{fig3}
\end{figure}

By bi-partitioning the whole graph $C$ in two successive steps as shown in
Fig.~$2$, we obtain a subgraph $B$ of $8$ nodes from the graph $C$ in the
first step, and in the second step we obtain a subgraph $A$ of $4$ nodes
from the subgraph $B$. The weights of the outgoing edges out of the subgraph
from a node are divided equally and assigned to other outgoing edges of the
node in the subgraph, such that the adjacency matrix of the subgraph remains
row-stochastic. For example, the weights of node $2$ in subgraph $A$ and
node $7$ in subgraph $B$ are changed in graph partitioning as shown in Fig.~$%
3$. With the above adjustment, the adjacency matrices of the subgraphs $B$
and $A$ are
\begin{equation}
S^{(B)}=%
\begin{pmatrix}
0 & \frac{1}{2} & \frac{1}{2} & 0 & 0 & 0 & 0 & 0 \\
0 & 0 & \frac{1}{2} & 0 & 0 & 0 & \frac{1}{2} & 0 \\
\frac{1}{2} & 0 & 0 & \frac{1}{2} & 0 & 0 & 0 & 0 \\
1 & 0 & 0 & 0 & 0 & 0 & 0 & 0 \\
0 & 0 & 0 & 0 & 0 & \frac{1}{2} & 0 & \frac{1}{2} \\
0 & 0 & 0 & 0 & 0 & 0 & 0 & 1 \\
\frac{1}{8} & \frac{1}{8} & \frac{1}{8} & \frac{1}{8} & \frac{1}{8} & \frac{1%
}{8} & \frac{1}{8} & \frac{1}{8} \\
0 & 0 & 0 & 0 & 0 & 0 & 1 & 0%
\end{pmatrix}%
,
\end{equation}%
and
\begin{equation}
S^{(A)}=%
\begin{pmatrix}
0 & \frac{1}{2} & \frac{1}{2} & 0 \\
0 & 0 & 1 & 0 \\
\frac{1}{2} & 0 & 0 & \frac{1}{2} \\
1 & 0 & 0 & 0%
\end{pmatrix}%
\end{equation}%
respectively. The Hamiltonian evolution path for obtaining the PageRank
vector of graph $C$ is $H_{A}\rightarrow H_{B}\rightarrow H_{C}$, and the
Hamiltonians $H_{A}$, $H_{B}$ and $H_{C}$ are constructed according to Eq.~($4$).

The algorithm requires $5$ qubits, we set the parameters $\beta =-1$
and $\omega =2$ in each step of the algorithm such that the resonance
condition is satisfied, and the coupling coefficient $c=0.002$, and perform
simulation of the mQRT algorithm. In the first step, the ground state $%
|\varphi ^{(A)}\rangle $ of $H_{A}$ can be easily calculated and prepared on
the register $R$ as the initial state of the algorithm
\begin{equation*}
|\varphi ^{(A)}\rangle =%
\begin{pmatrix}
0.62 & 0.34 & 0.62 & 0.34%
\end{pmatrix}%
^{T}.
\end{equation*}%
The ground state $|\varphi ^{(B)}\rangle $ of $H_{B}$ is obtained with
runtime $t_{1}=926$, in which the probe qubit reaches its maximum decay
probability, yielding the ground state
\begin{equation*}
|\varphi ^{(B)}\rangle =%
\begin{pmatrix}
0.54 & 0.33 & 0.46 & 0.29 & 0.10 & 0.14 & 0.45 & 0.26%
\end{pmatrix}%
^{T}
\end{equation*}%
of the Hamiltonian $H_{B}$. We perform the simulation of the algorithm for
the first step of the algorithm, and the results are shown in Fig.~$4$. We
can see that the resonance transition occurs at $\omega =2$ with almost unit
probability. In the second step, the ground state $|\varphi ^{(C)}\rangle $
of $H_{C}$ is obtained with runtime $t_{2}=917$, and
\begin{equation*}
|\varphi ^{(C)}\rangle \!=\!%
\begin{pmatrix}
\!0.50 & 0.26 & 0.52 & 0.27 & 0.05 & 0.07 & 0.27 & 0.13 & 0.19 & 0.24 & 0.10 & 0.08 & 0.05 & 0.27 & 0.15 & 0.19
\end{pmatrix}^{T},
\end{equation*}%
which has fidelity of $0.999$ with the ground state of the Google matrix of
the complete graph $C$. Here fidelity between two quantum states $|\varphi \rangle $ and $|\psi \rangle $ is defined as $F=\left\vert \langle \varphi
|\psi \rangle \right\vert ^{2}$, which measures the closeness of two states.
\begin{figure}[tbp]
\includegraphics[width=1\linewidth]{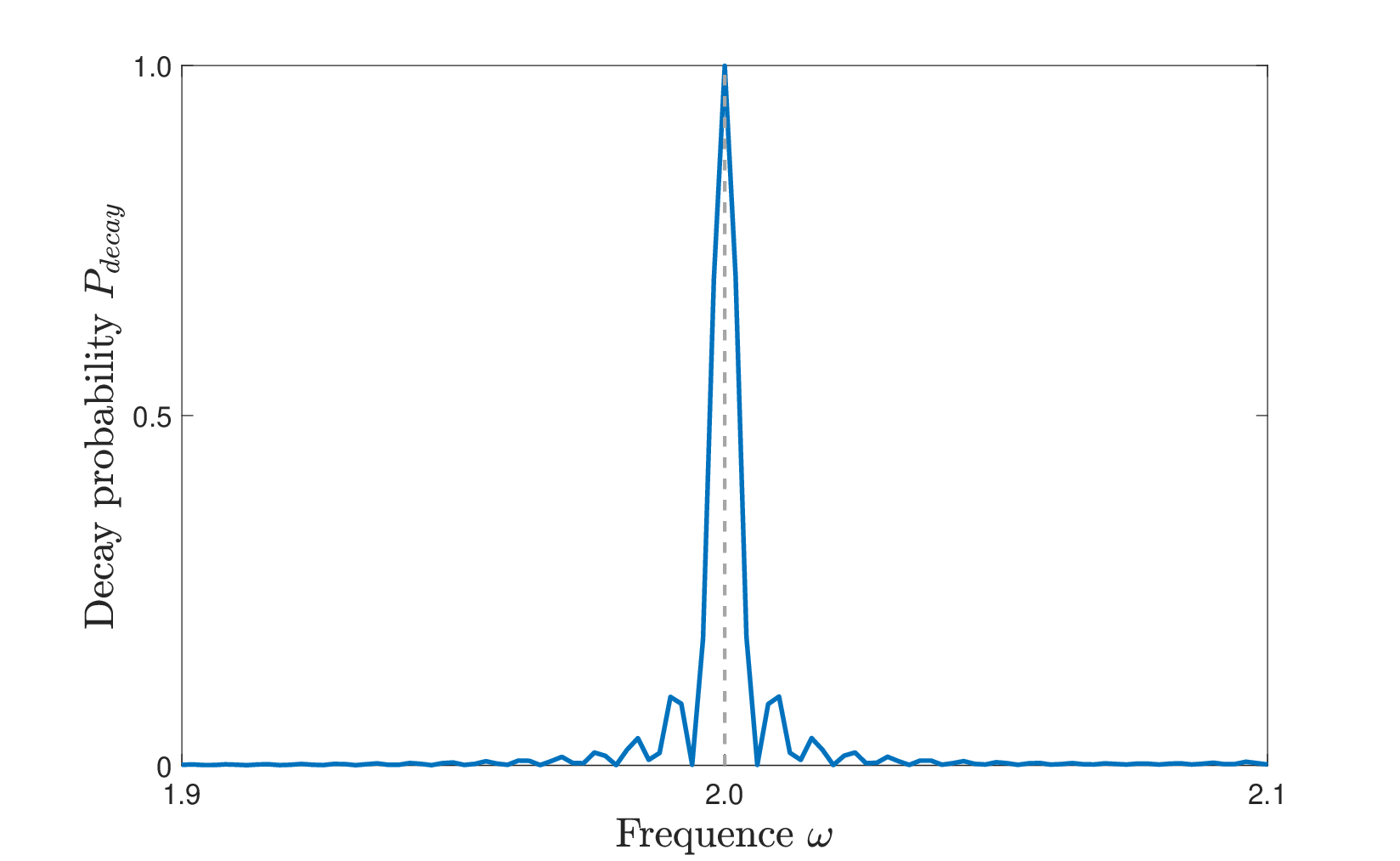}
\caption{(Color online.) Transition frequency spectrum between the ground
states of the Hamiltonian matrices $H_{A}$ and $H_{B}$ for the subgraphs $A$
and $B$ in Fig.~$3$. The blue solid curve represents the decay probability
of the probe qubit at different frequencies with the coupling coefficient $%
c=0.002$ and the evolution time $t=926$. The gray dotted vertical line
represents the transition frequency of $\protect\omega =2.0$.}
\label{fig4}
\end{figure}
\begin{figure}[tbp]
\makebox[\textwidth][c]{\includegraphics[width=1.2\linewidth]{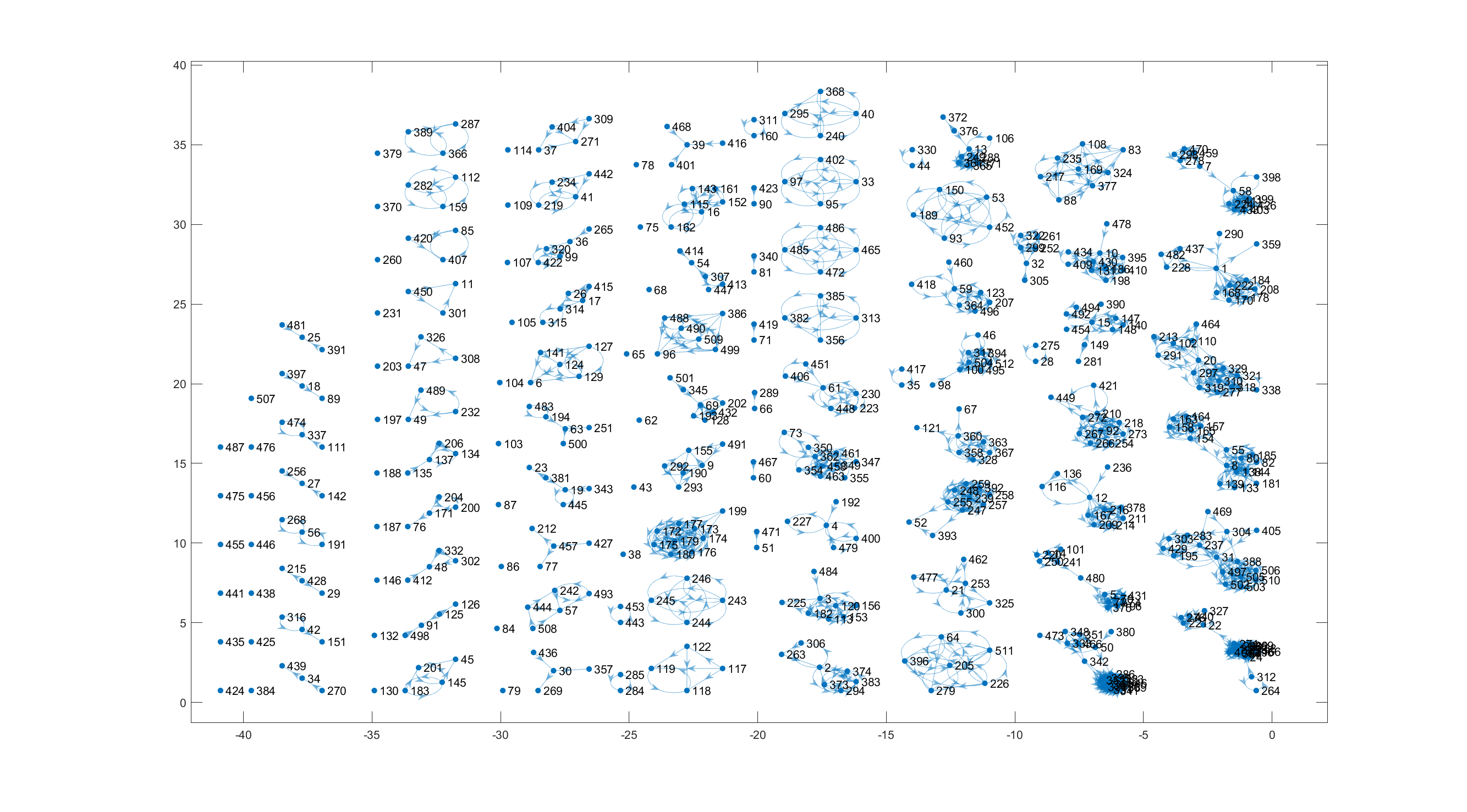}}
\caption{Graph of $512$ most active webpages of a web graph from Google.}
\label{fig5}
\end{figure}
\begin{figure}[tbp]
\makebox[\textwidth][c]{\includegraphics[width=1.2\linewidth]{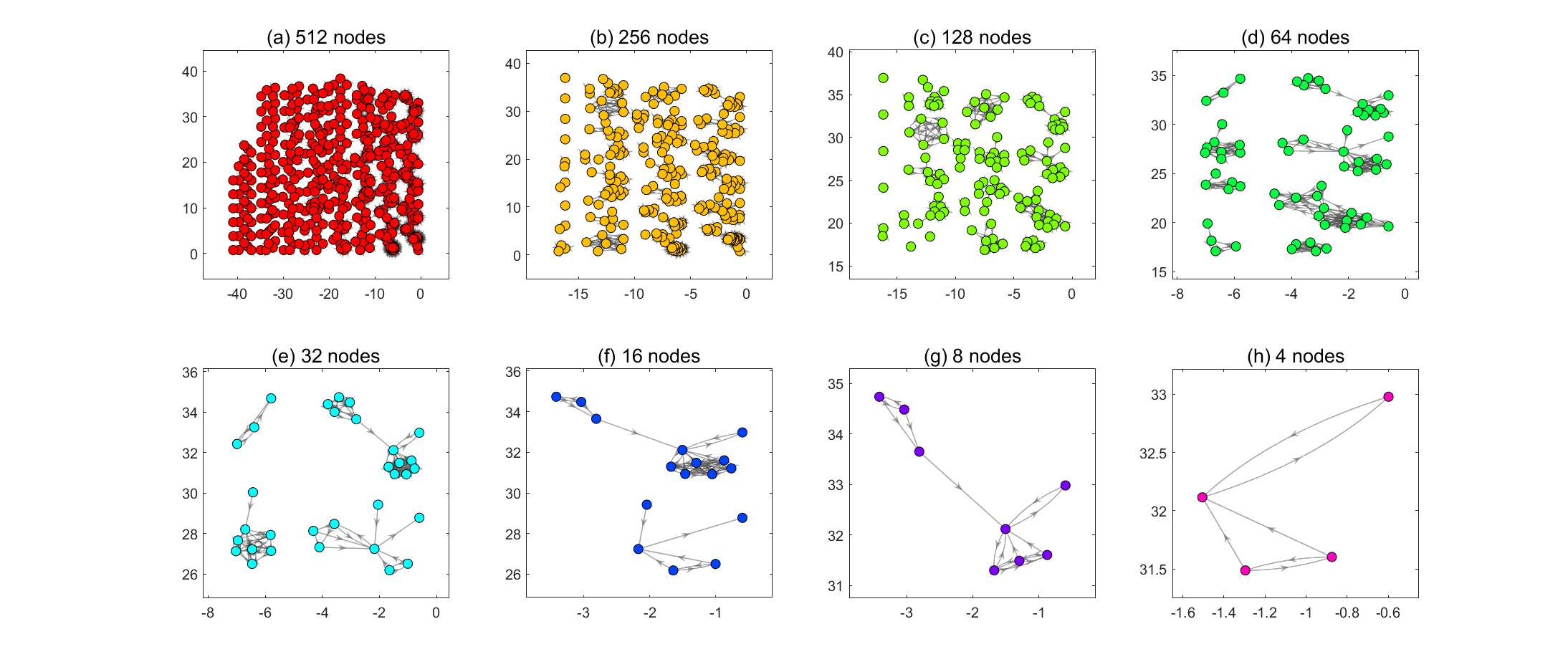}}
\caption{Subgraphs by bi-partitioning the graph of Fig.~$5$ in $7$
successive steps using the \textquotedblleft right-upper partitioning\textquotedblright\ method. Fig.~$6$(a) is the whole graph, and
Fig.~$6$(b) is the right half of Fig.~$6$(a), Fig.~$6$(c) is the upper half
of Fig.~$6$(b), and so on.}
\label{fig6}
\end{figure}
\begin{figure}[tbp]
\makebox[\textwidth][c]{\includegraphics[width=1.2\linewidth]{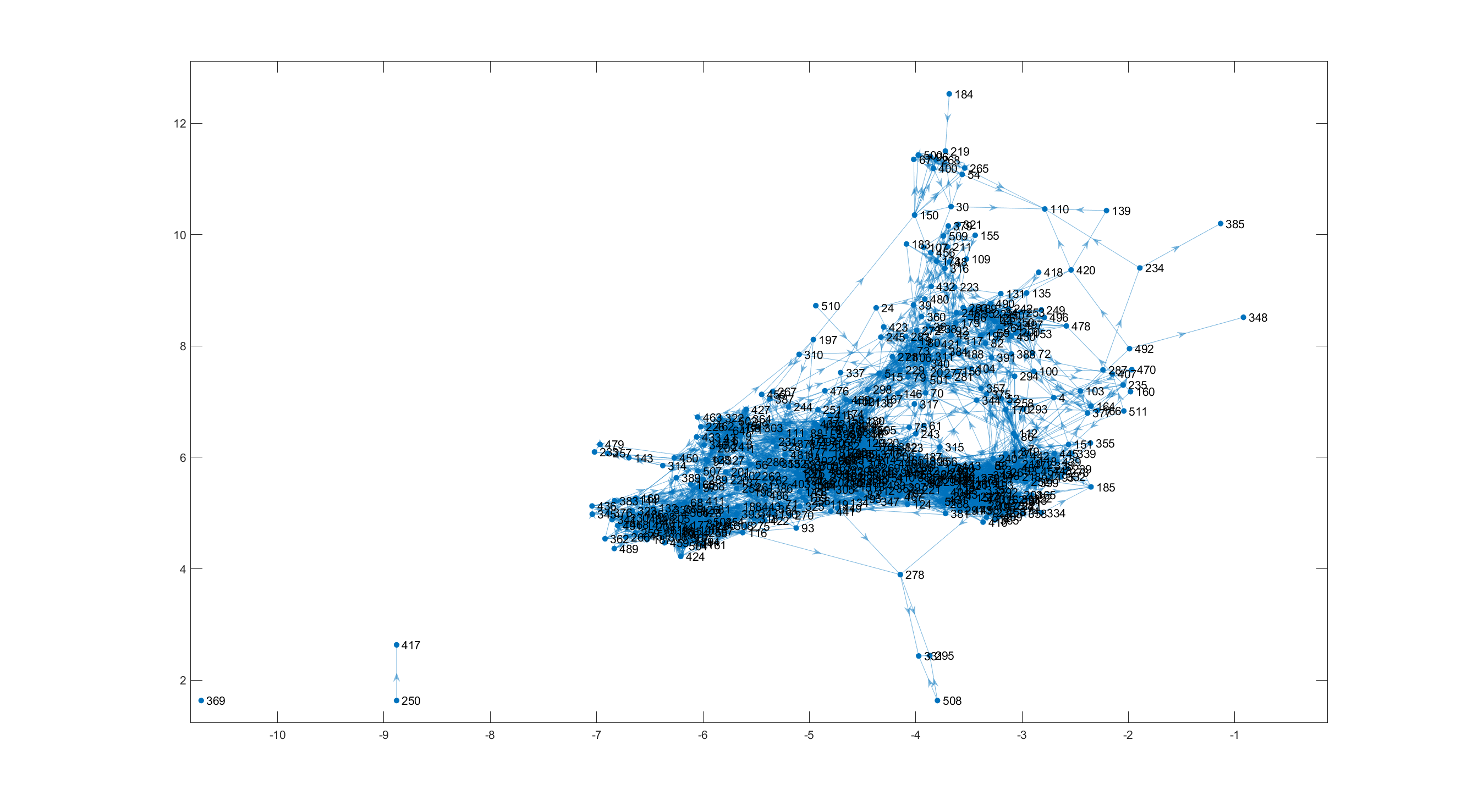}}
\caption{Graph of the $512$ most active webpages of a citation network.}
\label{fig7}
\end{figure}
\begin{figure}[tbp]
\makebox[\textwidth][c]{\includegraphics[width=1.2\linewidth]{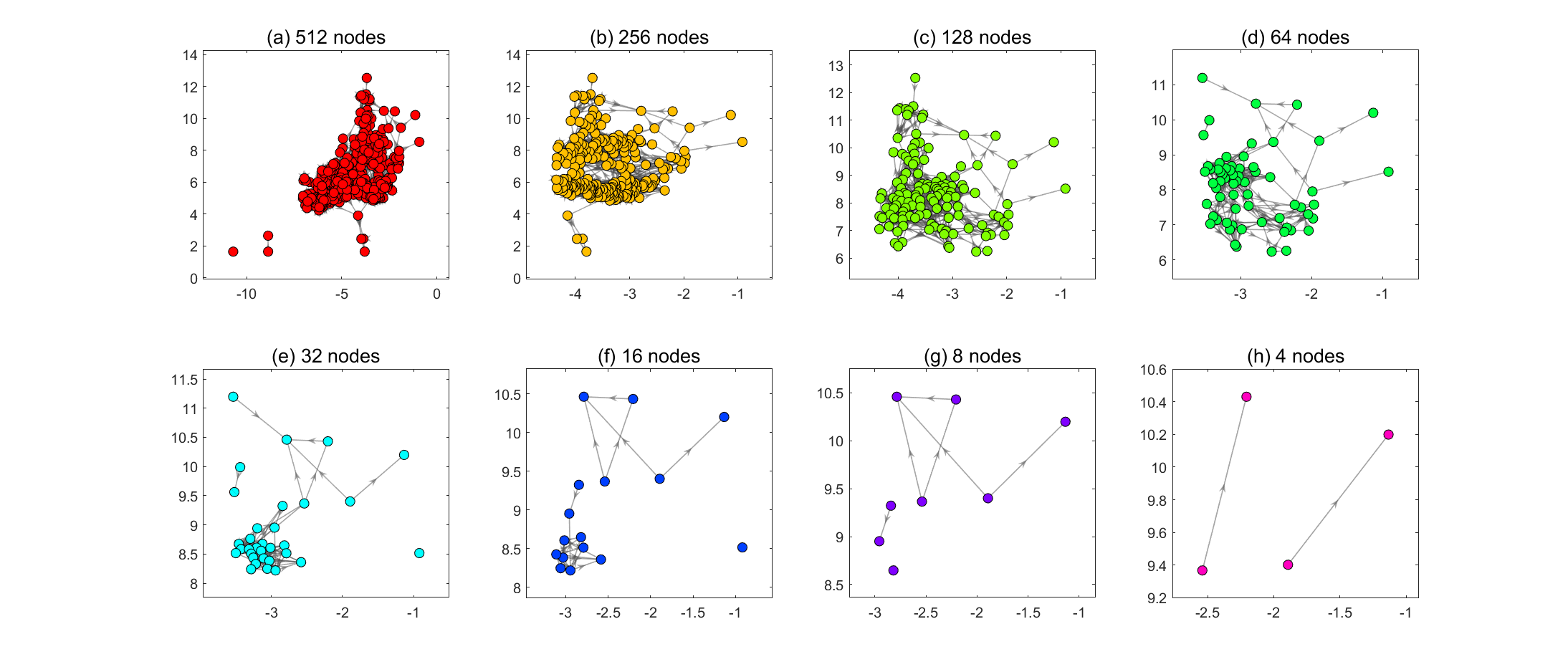}}
\caption{Subgraphs by bi-partitioning the graph of Fig.~$7$ in $7$
successive steps using the \textquotedblleft right-then-upper
partitioning\textquotedblright\ method as in that of Fig.~$6$.}
\label{fig8}
\end{figure}

\subsection{A Google web graph from SNAP}

In the following we use two practical networks as examples to verify the
effectiveness of the mQRT algorithm in obtaining the PageRank vector of a
Google matrix. We conduct systematic numerical simulations using the web
graph data provided by the Stanford Network Analysis Project~(SNAP) as a
benchmark, which completely records the real link topology of Google
webpages. The web graph contains $875,713$ nodes representing webpages and $%
5,105,039$ directed edges representing hyperlinks between webpages~\cite%
{Leskovec2014,webgraph}. We select the top $512$ nodes with the highest link
activity, and the network structure of the web graph is shown in Fig.~$5$.

For this directed graph of $512$ nodes, we apply a right-upper partitioning
strategy to divide the graph hierarchically: Vertically bisect the graph and
preserve the right subgraph, followed by horizontally bisecting the right
subgraph and retaining the upper part. In the first step, we divide the
graph evenly into left and right halves, respectively, and select the $256$
nodes that are connected closely on the right half side of the graph as
subgraph $1$; then, subgraph $1$ is divided into upper and lower halves, and
we select the $128$ nodes on the upper half of subgraph $1$ as subgraph $2$.
This process is repeated iteratively until termination conditions are met,
this forms a total of $7$ partitions as shown in Fig.~$6$. The Hamiltonian
for the whole graph is constructed as $H_{7}$ based on Eq.~($4$), whose
ground state $|\varphi ^{(7)}\rangle $ is the PageRank vector of the Google
matrix of the graph. Correspondingly, for the subgraph $k$, the Hamiltonian $%
H_{k}$ is constructed, and the ground state is $|\varphi ^{(k)}\rangle $.
Based on the above partition of the graph, the overlap between ground states
of two adjacent Hamiltonians are $d_{1}=0.32,$ $d_{2}=0.42,$ $d_{3}=0.46,$ $%
d_{4}=0.54,$ $d_{5}=0.39,d_{6}=0.45,d_{7}=0.49$, respectively. The partition
method of the graph is not unique. To verify the robustness of the
algorithm, we use a right-lower partitioning method for the graph of the
webpages. The overlaps remain polynomially large, and the final quantum
state fidelity consistently approaches $0.999$, confirming the robustness of
the algorithm.

\subsection{A graph of citation network from SNAP}

We also test our algorithm on a citation network, in which nodes represent
papers and edges represent citations. We use an arxiv high energy physics
paper citation network that contains $34,546$ nodes and $421,578$ edges as
an example~\cite{Leskovec,citation}. The web graph of the top $512$ most
cited papers are shown in Fig.~$7$. We still apply the right-upper
partitioning method to divide the graph in $7$ steps as shown in Fig.~$8$.
The overlaps between the PageRank vectors of Google matrices of two adjacent
subgraphs are $d_{1}=0.79$, $d_{2}=0.63$, $d_{3}=0.52$, $d_{4}=0.51$, $%
d_{5}=0.38$, $d_{6}=0.45$, and $d_{7}=0.47$, respectively. If the overlap
between the PageRank vectors $|\varphi _{0}^{(k)}\rangle $ and $|\varphi
_{0}^{(k+1)}\rangle $ of the subgraphs $D_{k}$ and $D_{k+1}$ is small, we
can use the graph $D_{k}^{\prime }=D_{k+1}-D_{k}$, i.e., the graph obtained
by removing all vertices of $D_{k}$ from $D_{k+1}$. If we take the
complementary subgraph of the graph in each step of the this example, the
corresponding overlaps are $d_{1}^{^{\prime }}=0.67$, $d_{2}^{^{\prime
}}=0.73$, $d_{3}^{^{\prime }}=0.46$, $d_{4}^{^{\prime }}=0.50$, $%
d_{5}^{^{\prime }}=0.44$, $d_{6}^{^{\prime }}=0.46$, and $d_{7}^{^{\prime
}}=0.45$, respectively. This shows the robustness of the algorithm.

\section{Discussion}

In the following we compare our algorithm to the other quantum algorithms in
obtaining the quantum state encoding the PageRank vector of the Google
matrix. In the QAE algorithm for solving this problem, the system is evolved
adiabatically from the ground state of an initial Hamiltonian to that of the
problem Hamiltonian. While our algorithm is digital, therefore it can be implemented on a circuit model and is compatible with error correction. It has the advantage that we only need to implement time-independent Hamiltonian evolution in each step which can be implemented efficiently through Hamiltonian simulation algorithms on a quantum computer. Compared to another digital quantum algorithm for solving the problem based on the HHL algorithm, our algorithm uses fewer resources, it requires only one ancillary qubit and the circuit is simple. While the algorithm based on the HHL algorithm requires a large number of ancillary qubits, and the circuit is more complicated than our algorithm. In our algorithm, the quantum state encoding the PageRank vector is obtained in $m$ steps, where $m$ scales as $O(\log N)$. There is cost of encoding the intermediate Hamiltonians and perform Hamiltonian simulation in each step,  and the cost is proportional to the number of steps $m$.

In conclusion, we present a quantum algorithm for obtaining the quantum
state encoding the PageRank vector of the Google matrix through multistep
quantum resonant transitions. In the algorithm, the graph of the Google
matrix is partitioned into a series of nested subgraphs with decreasing
sizes, a sequence of intermediate Hamiltonians are constructed based on the
subgraphs. The problem of solving the quantum state that encodes the
PageRank vector is transformed to finding the ground state of the problem
Hamiltonian associated with the Google matrix, and the state is obtained
step by step via the mQRT algorithm. The efficiency of the algorithm depends
on the energy gap between the ground and the first excited states of each
Hamiltonian that is always polynomial large, and the overlap between ground
states of two adjacent Hamiltonians. If by applying an appropriate
partitioning method as shown in this work, the overlap between the ground
states of two adjacent Hamiltonians is also polynomial large, then the mQRT
algorithm can be run efficiently. At the end of the algorithm, we obtain the quantum state that encodes the PageRank vector $|\pi _{0}^{\left( m\right)
}\rangle =\sum_{k=0}^{N-1}c_{k}|k\rangle $, where $\sum_{k=0}^{N-1}\left%
\vert c_{k}\right\vert ^{2}=1$, and the states $|k\rangle $ are the
computational basis state encoding the webpages. We need to measure the
quantum state in order to obtain the PageRank of the webpages. Usually we
are interested in the top-ranked webpages that correspond to large
components of the quantum state. By performing measurement on the state $%
|\pi _{0}^{\left( m\right) }\rangle $ in the computational basis states, the
probability of obtaining the state $|k\rangle $ is proportional to $%
\left\vert c_{k}\right\vert ^{2}$. Therefore, webpages with high rank are
sampled with high probability. The number of measurements $M$ needed to
estimate $\left\vert c_{k}\right\vert ^{2}$ is $M=O\left( \epsilon^{-2} \right)$ within an error $\epsilon $~\cite{VenegasAndraca2013b}.

\begin{acknowledgements}
This work was supported by the National Key Research and Development Program
of China~(2023YFA1009103) and the Natural Science Fundamental Research
Program of Shaanxi Province of China~(Grant No.~2022JM-021).
\end{acknowledgements}

\end{document}